# Measurement of the neutron flux at the Canfranc Underground Laboratory with HENSA

S E A Orrigo[1], J L Tain[1], N Mont-Geli[2], A Tarifeño-Saldivia[2], L M Fraile[3], M Grieger[4], J Agramunt[1], A Algora[1], D Bemmerer[4], F Calviño[2], G Cortés[2], A De Blas[2], I Dillmann[5], A Domínguez Bugarín[3], R García[2], E Nacher[1] and A Tolosa[1]

[1] Instituto de Física Corpuscular (IFIC), CSIC-Univ. Valencia, Spain
[2] Institute of Energy Technologies (INTE), Technical University of Catalonia (UPC), Barcelona, Spain
[3] Grupo de Física Nuclear & IPARCOS, Universidad Complutense de Madrid (UCM), Madrid, Spain
[4] Helmholtz-Zentrum Dresden-Rossendorf (HZDR), 01328 Dresden, Germany
[5] TRIUMF, 4004 Wesbrook Mall, Vancouver, British Columbia V6T 2A3, Canada

E-mail: Sonja.Orrigo@ific.uv.es

**Abstract**. We have performed a long-term measurement of the neutron flux with the High Efficiency Neutron Spectrometry Array HENSA in the Hall A of the Canfranc Underground Laboratory. The Hall A measurement campaign lasted from October 2019 to March 2021, demonstrating an excellent stability of the HENSA setup. Preliminary results on the neutron flux from this campaign are presented for the first time. In *Phase* 1 (113 live days) a total neutron flux of $1.66_2 \times 10^{-5}$ cm$^{-2}$ s$^{-1}$ is obtained. Our results are in good agreement with those from our previous shorter measurement where a reduced experimental setup was employed.

## 1. Introduction

An important limitation to rare event searches pursued in underground laboratories is originated by neutrons. Typically the neutron component arising from cosmic-ray muons is largely suppressed underground. Nevertheless, radiogenic neutrons are still produced in the rocks by ($\alpha$,n) reactions and spontaneous fission. Because of their large penetrability, neutrons can induce background signals in the detectors [1,2] affecting nuclear astrophysics, neutrino and dark matter experiments. On top of that, neutrons of different energies are expected to affect diversely the various low-rate experiments.

Therefore it is important to measure and characterize the neutron flux as a function of energy very close to the experiments. Aiming for this goal, we carried out a long-term measurement with the High Efficiency Neutron Spectrometry Array (HENSA [3]) in Hall A of the Canfranc Underground Laboratory (LSC) [4], which is located at 2500 m.w.e. [5] under Mount Tobazo in the Aragonese Pyrenees (Spain). The measurement in Hall A lasted from October 2019 to March 2021. The setup was operated remotely and monitored on a daily basis, showing it was remarkably stable. Calibrations with $^{252}$Cf neutron sources were performed in between the different measurement periods (*Phases*). Here we present results for the neutron flux from *Phase* 1 for the first time, covering a total of 113 live days.

## 2. The High Efficiency Neutron Spectrometry Array

Proportional counters filled by $^3$He gas are detectors characterized by a high efficiency for thermal neutrons, long-term stability and high background discrimination power [2,6-8]. Hence they are well suited to measure the neutron flux underground, which is also a low-rate measurement.

The HENSA spectrometer is an array of 10 cylindrical, 60 cm long proportional counters, filled by $^3$He gas ($P$ = 10 atm) and embedded in high-density polyethylene (PE) moderators of different thickness. HENSA, as mounted in the Hall A at LSC, is shown in figure 1. The working principle is the same as the Bonner Spheres Spectrometer (BSS) [6], with the advantage that the bigger size of the tubes allows HENSA to achieve a neutron response of a factor 10 higher than conventional BSS (5 to 15 times higher depending on the particular neutron energy considered, see figure 2). The different covering materials employed for the various counters are listed in table 1 and include PE moderators of different thickness as well as Cd (absorber for thermal neutrons) and Pb (converter and multiplier for high-energy neutrons). This choice of materials makes it possible to achieve sensitivity over a wide range of neutron energies, ranging from thermal to 10 GeV. The HENSA response to neutrons has been obtained by Monte Carlo simulations performed with both FLUKA [8] and GEANT4, with excellent agreement. Figure 2 shows the GEANT4 response matrix in 90 energy bins [9].

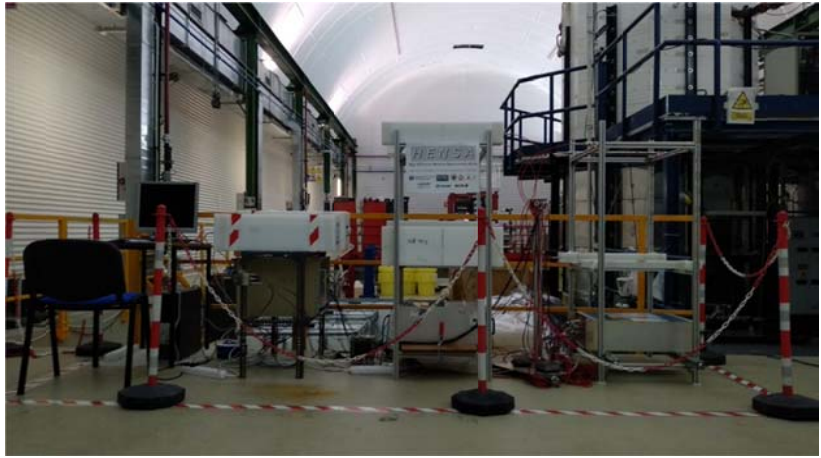

**Figure 1**. The High Efficiency Neutron Spectrometry Array HENSA at the LSC Hall A.

**Table 1.** Details of the shielding materials used in the various $^3$He detectors composing the HENSA spectrometer.

| Detector | Shielding |
|---:|---|
| 1 | Bare |
| 2 | PE  (4.5x4.5x70 cm$^3$) |
| 3 | PE   (7x7x70 cm$^3$) |
| 4 | PE  (12x12x70 cm$^3$) |
| 5 | PE  (18x18x70 cm$^3$) |
| 6 | PE  (22.5x22.5x70 cm$^3$) |
| 7 | PE  (27x27x70 cm$^3$) |
| 8 | PE  (21x21x70 cm$^3$) + Pb (5 mm) |
| 9 | Cd  (0.5 mm) |
| 10 | PE  (25x25x70 cm$^3$) + Pb (10 mm) + Cd (0.7 mm) |

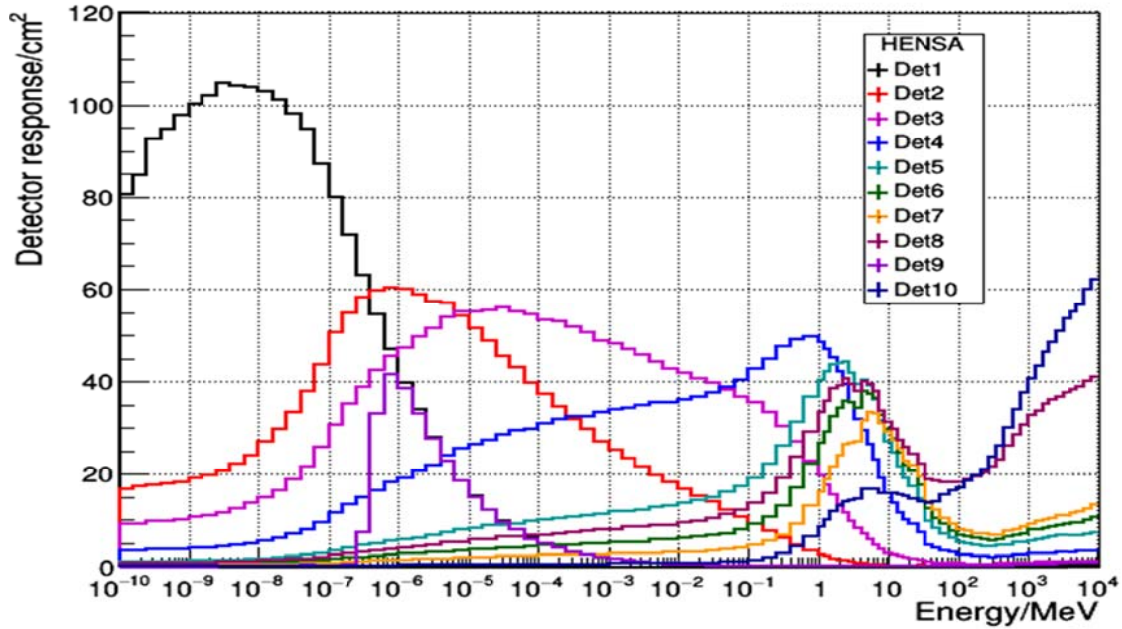

**Figure 2**. HENSA response to neutrons simulated with GEANT4 [9]. Lines of different colours correspond to the detectors listed in table 1.

## 3. Data analysis and preliminary results from *Phase* 1

Preliminary results on the first 113 live days of data (*Phase* 1) are presented in this section. Well-established methods are employed for the data analysis [2,7,8]. Thermal or moderated neutrons are detected indirectly in the $^3$He proportional counters through the $^3$He(n,p)$^3$H reaction releasing 764 keV. The p and $^3$H products share this available energy and can deposit it either fully or partially by leaving the detector before being fully stopped. This process produces a spectrum of deposited energy which is characteristic of neutron detection. An example of such a spectrum is shown in figure 3 for the thermal detector (bare counter), where the blue dash-filled line is the result of a calibration with a moderated $^{252}$Cf neutron source (*pure* neutron response) fitted to the experimental spectrum (red empty line) in the region of the 764 keV neutron peak. The linear α background originated from the counter walls (cyan dot-filled line) is also taken into account. With this procedure, the neutron rate $n_i$ for detector $i$ is determined from the corresponding energy spectrum.

The neutron rate observed in each detector results from a combined effect of the incoming neutron flux $\Phi_j$ and the detector response $\varepsilon_{ij}$:

$$n_i = \sum_j \varepsilon_{ij} \Phi_j \qquad (1)$$

Knowing the response matrix $\varepsilon_{ij}$ for each energy bin $j$, one is able to determine the neutron flux by using a deconvolution procedure based on iterative algorithms [10]. Parametric functions for thermal, epithermal, evaporation peak and high-energy region from [11] are used to define an initial neutron flux distribution (*prior* information) to feed the deconvolution algorithm more realistically than using a flat distribution. The energy spectrum of the neutron flux obtained using the expectation-maximization (EM) method, based on the Bayes theorem (see [10] for details), is shown in figure 4. By integrating on neutron energy, in *Phase* 1 a total neutron flux of $1.66_2 \times 10^{-5}$ cm$^{-2}$ s$^{-1}$ is obtained. Since the comparison of the results determined employing different deconvolution algorithms allows one to estimate the uncertainties inhering the deconvolution procedure [7,10], the determination of the neutron flux with other algorithms (such as, e.g., the maximum-entropy method [10]) is ongoing.

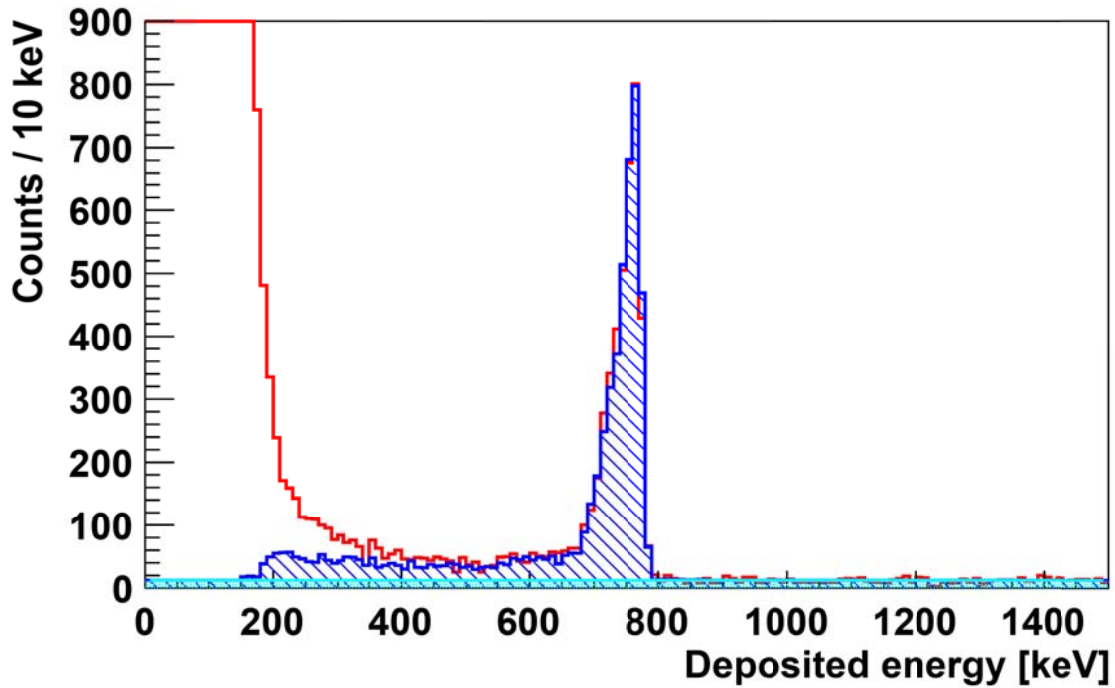

**Figure 3**. Spectrum measured in the $^3$He thermal detector (red line). The peak at low energy is due to electronic noise and γ-ray background, while the neutron-detection peak is visible at 764 keV. The blue dash-filled line is the *pure* neutron response from a measurement with $^{252}$Cf. The cyan dot-filled line is the linear background from the intrinsic α-radioactivity of the counter.

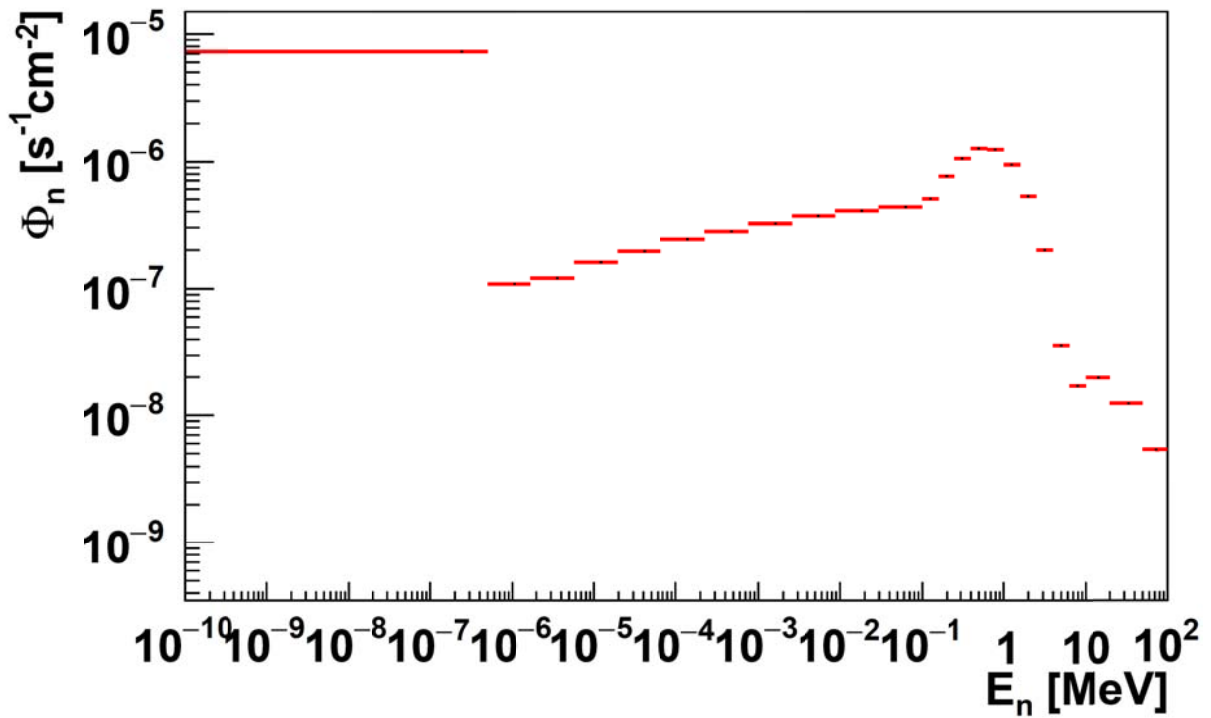

**Figure 4**. Energy distribution of the neutron flux obtained with the EM method, the response of figure 2 rebinned in 24 bins as in [7] and a *prior* distribution defined using functions from [11].

These results can be compared to those of our previous measurement done at the LSC Hall A in 2011 [2,7]. At that time Hall A was almost empty, the measurement lasted 26 days and the setup comprised 6 counters only, with a $^3$He pressure of 20 atm. Taking into account these differences, there is a very reasonable agreement of the present total neutron flux with the previously-measured value of $1.38_{14} \times 10^{-5}$ cm$^{-2}$ s$^{-1}$ [7]. The comparison of the neutron rates measured in the detectors that were used in both measurements, shown in table 2, is also satisfactory considering the different pressure of the $^3$He tubes in the two cases. Moreover, the comparison of the energy spectra in the two measurements (see, e.g., figure 3 here and figure 4 in [7]) shows that the background observed in [2,7] at low energy is now considerably suppressed, likely thanks to the change of detector pressure pointing to γ-ray sensitivity as the origin. The electronic noise has also been reduced by improving the electromagnetic compatibility of our electronic setup. Finally, the addition of other 4 proportional counters to the setup (detectors number 1, 8, 9 and 10 in table 1) allowed us to achieve an increased sensitivity at both extremes of the neutron energy range (see figure 2).

The data analysis of *Phase*s 2 and 3 is in progress as well as the study of the long-term evolution of the neutron rate. Moreover, a new long-term measurement started in March 2021 in Hall B of LSC, located close to the ANAIS-112 dark matter experiment [12], and is currently running [13].

**Table 2.** Comparison of the neutron rates from the present (*Phase* 1) and previous measurements, for the detectors used in both. Detectors are labelled as in table 1.

| Detector | Neutron rate [$\times 10^{-4}$ s$^{-1}$] | Neutron rate [$\times 10^{-4}$ s$^{-1}$] from [7] |
|---|---|---|
| 2 | 4.52(14) | 4.38(20) |
| 3 | 4.98(11) | 5.04(21) |
| 4 | 4.58(11) | 3.79(19) |
| 5 | 2.34(8) | 2.33(16) |
| 6 | 1.38(9) | 1.28(12) |
| 7 | 0.81(5) | 0.77(10) |

**Acknowledgments**
This work was supported by the Spanish MICINN Grants No. PID2019-104714GB-C21 and No. FPA2017-83946-C2-1-P (MCIU/AEI/FEDER). We acknowledge the support of the Generalitat Valenciana Grant No. PROMETEO/2019/007. We are grateful to Laboratorio Subterráneo de Canfranc for hosting the HENSA spectrometer and for the support received by the LSC personnel during the measurement campaign.